\documentclass{IEEEtran}
\usepackage{cite}
\usepackage{amsmath,amssymb,amsfonts}
\usepackage{graphicx}
\usepackage{hyperref}
\usepackage{xcolor}

\usepackage{multirow}
\graphicspath{{./eps/}}
\DeclareGraphicsExtensions{.eps}
\ifCLASSOPTIONcompsoc
  \usepackage[caption=false,font=normalsize,labelfont=sf,textfont=sf]{subfig}
\else
  \usepackage[caption=false,font=footnotesize]{subfig}
\fi
\def\BibTeX{{\rm B\kern-.05em{\sc i\kern-.025em b}\kern-.08em
T\kern-.1667em\lower.7ex\hbox{E}\kern-.125emX}}

\begin{document}
\title{Characterization of CMOS SPADs for future RICH Detectors}
\author{R.~Dolenec\thanks{*R.~Dolenec and H.K.~Yildirim contributed equally to this work.}*, H.K.~Yildirim*, G.V.~Tran, A.~Domenech, B.C.~Efe, W.Y.~Ha, U.~Karaca, P.~Singh, G.G.~Taylor, S.~Korpar, P.~Kri\v{z}an, R.~Pestotnik, A.~Seljak, E.~Charbon, C.~Bruschini  
\thanks{R.~Dolenec and P.~Kri\v{z}an are with the Faculty of Mathematics and Physics, University of Ljubljana, Slovenia and the Jo\v{z}ef Stefan Institute, Ljubljana, Slovenia. C.~Bruschini, E.~Charbon, W.Y.~Ha, P.~Singh, H.K.~Yildirim, G.V.~Tran, A.~Domenech, B.C.~Efe, U.~Karaca and G.G.~Taylor are with the AQUA Laboratory, École polytechnique fédérale de Lausanne (EPFL), Neuchâtel, Switzerland. S.~Korpar is with the Faculty of Chemistry and Chemical Engineering, University of Maribor, Slovenia and the Jo\v{z}ef Stefan Institute, Ljubljana, Slovenia. R.~Pestotnik and A. Seljak are with the Jo\v{z}ef Stefan Institute, Ljubljana, Slovenia.

This project has received funding from the Slovenian Research and Innovation Agency (project J1-50009) and the Swiss National Science Foundation (project No 200021E\_218853).

Corresponding authors: Rok Dolenec (rok.dolenec@ijs.si) and Claudio Bruschini (claudio.bruschini@epfl.ch).}
}

\maketitle

\begin{abstract}
In the planned or considered upgrades of LHCb, ALICE and Belle II experiments, the Ring imaging Cherenkov (RICH) detectors will have to be improved in order to function at increased beam interaction density. The photodetectors used in future RICH detector will have to provide high granularity, single photon sensitivity and excellent timing, while being exposed to a couple of \mbox{10$^{13}$ 1-MeV neutron equivalent/cm$^2$} of background irradiation during total experiment run time. The spadRICH project is developing a CMOS single-photon avalanche diode (SPAD) based photodetector specifically optimized for the application of the planned RICH detectors, which includes neutron radiation hardness and cryogenic operation. In this work we present recent experimental characterization studies of existing SPADs produced in 55 nm BCD and 110 nm CMOS image sensor technologies. Main results include dark count rate (DCR) measurements with SPADs irradiated up to \mbox{10$^{12}$ 1-MeV neutron equivalent/cm$^2$} and cooled down to liquid nitrogen temperature.
\end{abstract}

\vspace{-0.6cm}
%
\section{Introduction}

The planned upgrades of Ring Imaging Cherenkov (RICH) detectors in experiments such as LHCb, ALICE and Belle II will require photodetectors meeting very high performance requirements. Silicon photomultipliers (SiPMs) are a promising candidate photodetector, with the main remaining challenge being their sensitivity to neutron radiation. At the expected irradiation levels of a couple of \mbox{10$^{13}$ 1-MeV neutron equivalent/cm$^2$} (n$_{eq}$/cm$^2$), the SiPMs would need to be operated close to liquid nitrogen temperature in order to keep sufficient performance until the end of experiment \cite{Consuegra2024}.

The spadRICH project is developing a digital SiPM \cite{Muntean2017, Muntean2020} optimized specifically for the demands of future RICH detectors. We plan to achieve radiation hardness by implementing radiation hard design techniques at transistor and single photon avalanche diode (SPAD) level, integrate electronics with functionality to mitigate radiation damage, and use microlens arrays to reduce the sensitive SPAD volume. Here we present the characterization experiments performed with existing SPAD samples, focusing on cryogenic operation and effects of neutron irradiation.
%
\begin{figure*}[hbt]
\centering
\includegraphics[width=0.75\textwidth]{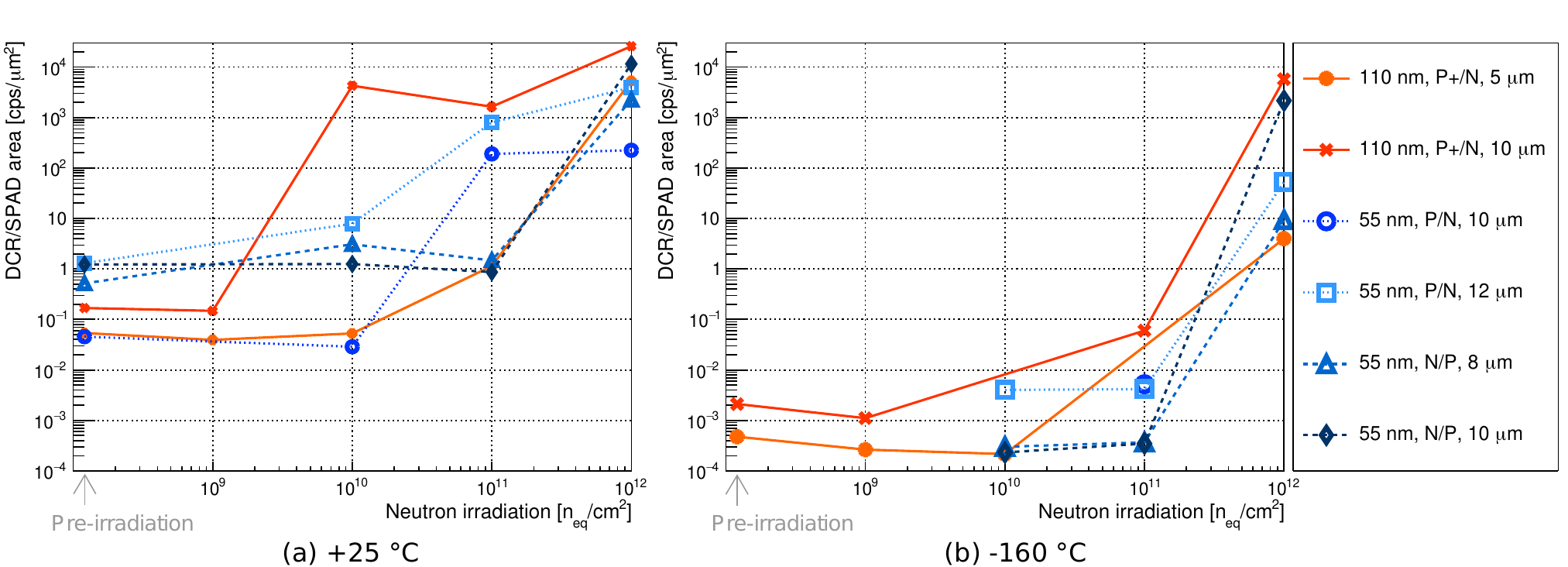}
\caption{Active area normalized DCR vs. neutron fluence measured with different SPADs (legend listing: technology node, junction type, SPAD diameter) at RT (a) and close to LN (b), all at 6V excess bias. The data at different irradiation levels were obtained with different samples of the same SPAD type. Some data points are missing at low temperatures due to disconnects of power or signal lines during continuous temperature cycling.}
\label{figResCryo}
\end{figure*}

\begin{figure*}[hbt]
\centering
\includegraphics[width=0.775\textwidth]{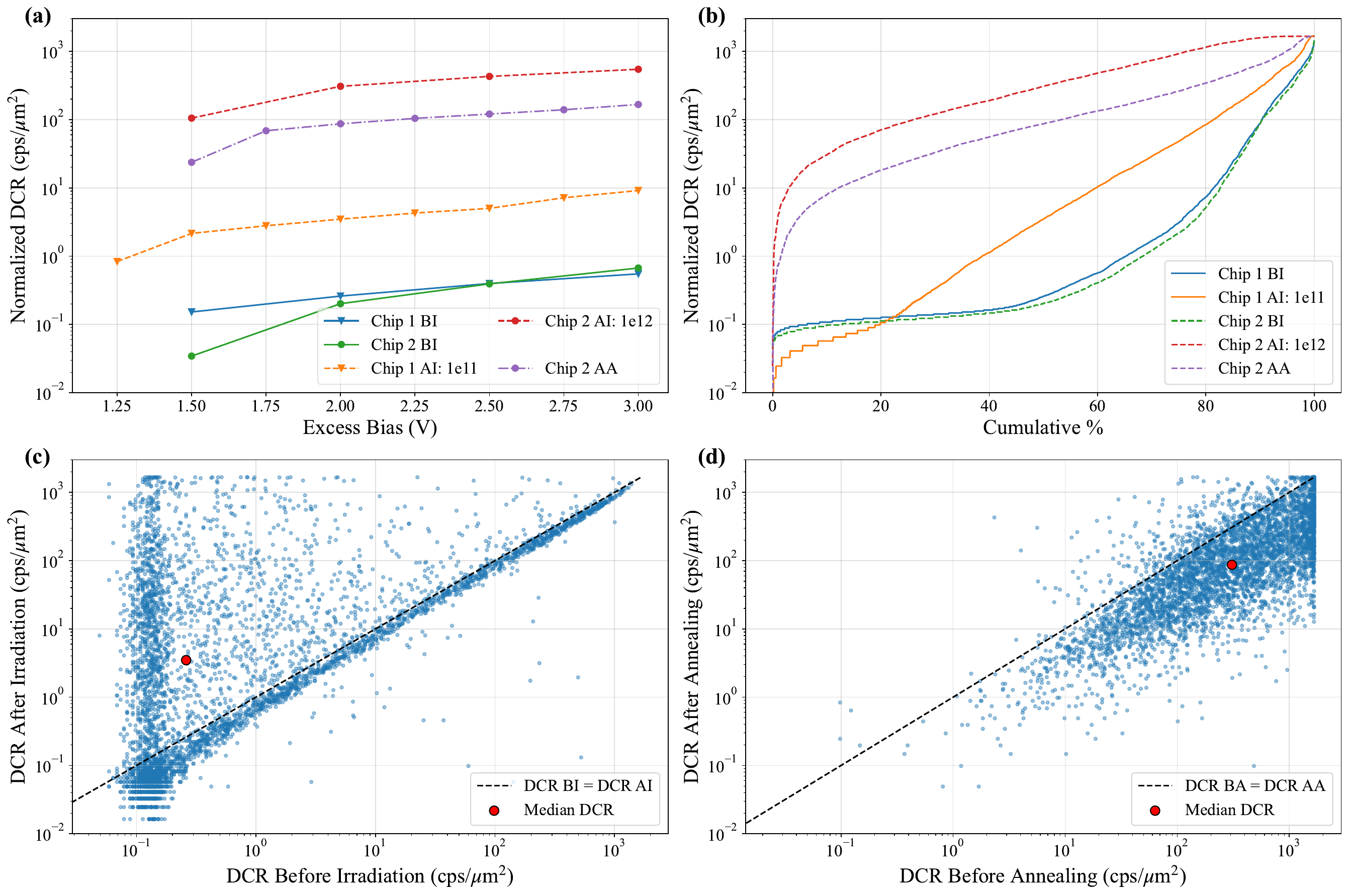}
\caption{Active area normalized room temperature DCR measured for a 144$\times$32 SPAD array: data from Chip 1~and Chip~2 irradiated with neutrons at \mbox{10$^{11}$ n$_{eq}$/cm$^2$} and \mbox{10$^{12}$ n$_{eq}$/cm$^2$}, respectively, Chip~2 further annealed at high temperature (legend: BI: before irradiation, AI: after irradiation, AA: after annealing). (a) Median DCR vs excess bias. (b) DCR distribution curves at 2V excess bias. (c) Scatter plot showing normalized pixel DCR before vs after irradiation for Chip~1. (d) Scatter plot showing normalized pixel DCR before vs after annealing for Chip~2.}
\label{figarray}
\end{figure*}

\section{Materials and methods}

Several individual SPADs designed in 55~nm BCD technology \cite{SPAD55, SPAD55b, SPAD55c} and 110~nm CMOS image sensor technology \cite{SPAD110} were characterized at temperatures between room (RT) and liquid nitrogen (LN) temperature, before and after irradiation with neutrons up to \mbox{10$^{12}$ n$_{eq}$/cm$^2$}. Besides CMOS technology, the tested SPADs differed also in junction designs and diameters. The characterization setup, based on the one presented in \cite{Consuegra2024, Prev55110,PW25}, enabled SPAD samples to be stabilized at selected temperatures between RT and LN. 
The DCR measurement was performed in two regimes: high rate ($>$~100~cps, using scaler) and low rate ($<$~100~cps, using custom TDC).

A 144$\times$32 SPAD array designed in a 110~nm CIS technology (same as mentioned above for individual SPADs) was also characterized at room temperature to obtain statistics of the device response to irradiation. The array contains 5~$\mu$m active diameter P+/N SPADs at a pitch of 50~$\mu$m. A binary frame front-end and readout architecture based on \cite{PhoenixOpEx} was implemented. Binary frames at 32 kHz in passive quenching mode were accumulated to obtain DCR maps.

\section{Results}

The DCRs measured with individual SPAD samples at different neutron irradiation levels and two temperatures are shown in Figure~\ref{figResCryo}. At RT, the DCR increased by about 4 orders of magnitude by irradiation of \mbox{10$^{12}$ n$_{eq}$/cm$^2$}. Cooling to $-$160~$^{\circ}$C recovers most of the DCR performance except for a couple of SPADs which seem to have received especially high damage. The samples produced in 55~nm BCD technology coped with irradiation slightly better than 110~nm CMOS SPADs, when taking their diameter into account. The results are in general consistent with the expectation that larger active area SPADs will be more susceptible to neutron damage.

Figure~\ref{figarray} shows DCR measurements from two of the 144$\times$32 SPAD arrays, labelled Chip~1 and 2, irradiated by \mbox{10$^{11}$ n$_{eq}$/cm$^2$} and \mbox{10$^{12}$ n$_{eq}$/cm$^2$} neutrons respectively. Chip~2 was further annealed at high temperature from 100$^{\circ}$ to 160$^{\circ}$ in 20$^{\circ}$ steps\cite{Wu2022}. Fig~\ref{figarray}(a) shows the median DCR of the array versus excess bias, showing 10$\times$ and 1500$\times$ increase in DCR at different doses, and a 3$\times$ decrease on Chip~2 after annealing. Fig~\ref{figarray}(b) shows the DCR distribution curves of the same measurements at 2V excess bias. Fig~\ref{figarray}(c) and (d) show scatter plots of the DCR of all pixels after radiation vs before radiation of Chip~1, and after annealing vs before annealing of Chip~2, respectively, at 2V excess bias.
\vspace{-0.4cm}
\section{Discussion}

Characterization experiments of several BCD and CMOS SPADs were conducted, with emphasis on cryogenic operation and neutron irradiation effects. While a SPAD can be badly damaged by neutrons and become noisy to the level of 10~kcps/$\mu$m$^2$, cooling close to liquid nitrogen temperature mitigates some of the damage. 
Data from SPADs across different technologies and configurations will guide the design of radiation-hard digital SiPMs optimized for cryogenic operation.
\vspace{-0.6cm}


\begin{thebibliography}{1}



\bibitem{Consuegra2024} D. Consuegra Rodríguez et al., 
Eur. Phys. J. C (2024) 84:970. 

\bibitem{Muntean2017} A. Muntean et al., 
2017 IEEE NSS/MIC Conference Record. 

\bibitem{Muntean2020} A. Muntean et al., 
IEEE Trans. Radiat. Plasma Med. Sci, 5(5) 671-678 (2020). 

\bibitem{SPAD55} F. Gramuglia et al., 
IEEE J. Sel. Top. Quantum Electron. 28(2) (2022) 3802410. 

\bibitem{SPAD55b} A. Morelle et al., 
Quantum Information and Measurements VI (2021). 

\bibitem{SPAD55c} F. Liu et al., 
IEEE J. Sel. Top. Quantum Electron. 30(1) (2024) 3801407. 


\bibitem{SPAD110} U. Karaca, 
Ph.D. dissertation, EPFL, 2024.



\bibitem{PW25} G.G. Taylor et al., 
Proc. SPIE (2025) 13373. 

\bibitem{Prev55110} R. Dolenec et al., 
JINST 20 (2025) P06052. 

\bibitem{PhoenixOpEx} H.K. Yildirim et al., 
Opt. Express. 34(3) (2026) 5064:5078. 

\bibitem{Wu2022} M.-Lo. Wu et al., 
Sensors. 22(8) (2022) 2919. 

\end{thebibliography}
\end{document}